\documentclass[twocolumn,amssymb,10pt,showpacs,prd]{revtex4-1}

\usepackage{graphicx}
\usepackage{amsmath}
\usepackage{bm}
\begin{document}
\title{ Role of Equation of State in formation of Black hole  } 

\author{Sanjay Sarwe}
\email{Electronics address:sbsarwe@gmail.com} 
\affiliation{Department of Mathematics, \\
S. F. S. College, Seminary Hill, Nagpur-440 006, India}

\begin{abstract} 
{We study the physical process of  gravitational collapse 
of a perfect fluid with a linear isentropic equation of state 
$p = k \rho, \ -1/3 < k \leq 1$. We consider a model with ansatz $v'(t,r)/v(t,r) = \xi'(r)$ 
that give rise to a family 
of solutions to Einstein equations with equation of state. 
The solution so obtained lead to homogeneous collapse that evolves from a regular initial 
data and the positivity of energy conditions. 
The collapse terminates in the formation of black hole.   
We show that as the parameter $k \rightarrow 1$, the formation of black hole
gets accelerated in time, revealing the significance of equation of state in black hole formation.}
\end{abstract}  
\maketitle

%\date{19$^{th}$ Feb 2015 }

\section{Introduction}

When the star heavier than  few solar masses has exhausted its internal nuclear fuel 
that supplies the outward pressure due to gas and radiation against the inward pulling gravitational force,
then the star cools down - it loses its equilibrium, the unbalanced gravity forces it to shrink,
and the perpetual gravitational contraction begins.
The end state of such contraction develops a space-time singularity for a wide range of
physically reasonable initial data 			
\cite{PSJ}.
Concerning such a space-time singularity, Penrose formulated the cosmic censorship conjecture (CCC) , 
	it states that '{\it a singularity of gravitational collapse of a massive star developed 
	from a regular initial 
	surface must always be hidden behind the event horizon of the gravity}'
	 \cite{rp}.
	This conjecture advocates the formation of black hole (BH) only
	 as against the formation of naked singularity (NS) 
	(wherein the time of formation of singularity precedes the epoch of formation of trapped surfaces).
	The CCC is fundamental to the well developed theory and astrophysical 
applications of black hole physics today. 

Oppenheimer and Snyder studied the spherically symmetric model of a homogeneous 
dust cloud that led to the general concept of trapped surfaces and the formation of BH
 \cite{OpSn}.
 The formation of the event horizon takes place here,
well in advance to the epoch of the formation of the
spacetime singularity, and hence it is necessarily hidden
behind the event horizon of the gravity, thus forming the BH.
This model invigorated the life of BH physics but it lacks pressure
which is a main constituent in the study of gravitational collapse.

It is known that the physical attributes of the matter field constituting a star 
are described by an equation of state relating the pressure and density of the matter field,
and hence it is important to know if the BH would form for an assumed equation
of state for the collapsing cloud.  The collapsing massive star evolves to having 
super dense states of matter close to 
the end stages of the collapse where the physical 
region has ultra-high densities, energies and pressures. 
Such an ultra-high density region of collapse needs to be 
described by a physically 
realistic equation of state but as of now such an equation of state is not 
precisely known, and also whether 
the equation of state would remain unchanged or  would it
actually evolve and change as the collapse develop? ,  these are some of the intriguing questions
\cite{SSJ}.

To describe the 
collapse of a massive star, we can choose the equation of 
state to be linear isentropic or polytropic after it loses its equilibrium configuration.
The gravitational collapse of a perfect fluid with a linear 
equation of state is of interest from both theoretical as well
as numerical relativity perspectives.

Self-similar perfect fluid collapse models with a linear 
equation of state were considered through numerical simulations 
by Ori and Piran
\cite{op}
and analytically by Joshi and Dwivedi 
\cite{dj2}
to show how black holes and naked singularities develop
as collapse final states in this scenario. 
Further, Goswami and Joshi
studied the case of an isentropic perfect fluid 
with a linear equation of state 
without the self-similarity assumption, wherein they
showed that the occurrence of BH and NS evolving from regular
initial data depends on the choice of rest of the free 
functions available
\cite{rgj}.

R. Goswami and P. Joshi have studied a special class of perfect fluid collapse models (wherein 
the mass function is assumed to be separable in terms of 
the physical radius of the cloud and
the time coordinate) 
which generalizes the homogeneous
dust collapse solution in order to include non-zero pressures and inhomogeneities into evolution 
\cite{RGPJ}.
It is shown that a BH is necessarily generated as end product of continued gravitational
collapse.

Our motivation also comes from certain other questions 
such as, what if the value of $k$
increases in the range $-1/3 < k \leq 1$ when a BH
 appears as collapse final state for a given 
spacetime dimension, will the BH formed sustain its nature?
If, it is so, will the formation of BH precede in time as $ k \rightarrow 1$ ?  
 We believe
 answers to these and similar issues would be important 
to understand better the physical aspects 
and the role of an equation of state in gravitational collapse of a star.

Therefore, our purpose here is to examine our motivation for a special class of solutions of
the Einstein field equations for a spherically symmetric  gravitational collapse of perfect
fluid to know explicitly how a homogeneous
density profile should behave in the later stages of collapse
and near the singularity as $ k \rightarrow 1$, 
so that the collapse end state would always be a BH.

In section II, the collapse with linear equation of state 
is studied through the special solution obtained that terminates into BH.
The formation of apparent horizon is investigated for parameter $k \rightarrow 1$ in section III.
The star model is completed by studying matching conditions in section IV.
The conclusions and remarks are specified in section V.

\section{Homogeneous Collapse with linear equation of state}
In comoving coordinates \ $(t, r, \theta, \phi)$, the general spherically symmetric metric,  
\begin{equation}
 ds^2 = - e^{2 \nu(t,r)}dt^2 +  e^{2 \psi(t,r)}dr^2 + R^2(t,r) d\Omega^2 \label {Dm01}
\end{equation}
describes spacetime geometry of a collapsing cloud
where \ $  d\Omega^2= d \theta^2 + \sin^2{\theta} d \phi^2  $ \ \
is the metric on a two-sphere.
The stress energy-momentum tensor for the type I matter fields
for perfect fluids is expressed  by
\[
 T^{\nu}_{\mu}  =  diag[-\rho, p, p, p ] .
\] 
Herein the quantities $\rho$ and $p$ are the physical entities representing energy
density and pressure respectively. 
The weak energy condition is a requirement that for every future directed time-like 
vector field $u^{\mu}$ and the stress-energy tensor $T_{\mu \nu}$ satisfies the relation 
$  T_{\mu \nu} u^{\mu} u^{\nu} \geq 0$ which implies 
${{\rho} \ {\geq}}0$; ${\rho+p \ {\geq}} 0$. 
Clouds perfect fluid relation is described through the linear equation of state
\begin{equation}
p(t,r) = k \ \rho(t,r) \ \text{where} \  k \in (- 1/3, 1] . \label{Des04}
\end{equation}
The Einstein field equations for the metric 
(\ref {Dm01}) are written as $( 8 \pi \texttt{G} =c=1)$
 \cite{rgj}
\begin{eqnarray}
  && \rho   = \frac{{F}'}{ R^{2} R'} \;  \;
  = - \frac{1} {k} \frac{\dot{{F}}}{R^{2} \dot{R}} \label {Dsfe05} \\
  && \nu \;'   = - \frac{k}{k + 1} [ \ln(\rho)]' \label {Dsfe06} \\
  && R' \dot{G} - 2 \dot{R} G \ {\nu}'  = 0  \label {Dsfe07} \\
  && F = R \left[ 1 - G + H \right] \; \;  \label {Dsfe08} 
\end{eqnarray}
	where the functions $G$ and $H$ are defined as 
$G(t, r) = e^{-2 \psi} {R'}^2$ and $H(t, r) = e^{-2 \nu} \dot{R}^2$
and the arbitrary function $F(t, r)$ has an interpretation of the mass function for the star. 
On any spacelike hypersurface $t = const.$, $F(t, r)$ determines the total mass of the star 
in a shell of comoving radius $r$. The weak energy conditions 
restricts  $F$, namely by $F(t, r) \geq 0$, and we have $F(t, 0) = 0$
to preserve the regularity of the model at all the epochs.

We introduce a new function  $v(t, r)$ by
$v(t, r)= R/r$,  
and using the scaling independence of the comoving
coordinate $r$, we write 
\cite{dj2}
\begin{equation}
R(t, r) = r \; v(t, r) .\label{Df09}
\end{equation}
In the continual collapse of the star,  we have 
$\dot{R} < 0$, it specifies that the physical radius $R$ of the collapsing cloud
keeps decreasing in time and ultimately, it reaches $R = 0$, and it denotes 
spacetime singularity, namely the  
 shell-focusing singularity at $R = 0$, where all the matter shells collapse to a
vanishing physical radius
at the epoch $t = t_s$.

The mass function $F(t, r)$ acts 
suitably at the regular center so that the density remains finite
and regular there at all times till the occurrence 
of singular epoch.  The Misner-Sharp mass function for the cloud can be written in general 
as,
\begin{equation}
{F}(t,r) = r^{3} M(r,v) \label {Dsm12}
\end{equation}
where the function $M(r,v)$ is regular and continuously twice
differentiable. On using equation (\ref {Dsm12}) in equation (\ref
{Dsfe05}), we have
\begin{equation}
\rho =  \frac{3 M + r [
M_{, r} + M_{, v}
 \ v']}{ v^{2} ( v + r v')}  
  =   - \frac{ M_{, v}} { k \ v^{2}}.  \label {Dsm13}
\end{equation}
 
We rearrange terms in equation (\ref{Dsm13}) and express it as
\begin{equation}
 k \; r {M}_{,\ r} +[(k + 1) r v' + v]
{M}_{,\ v} = - 3 k M \; .  \label {Dsv14.1}
\end{equation}

Next to obtain the general solution of equation (\ref {Dsv14.1}),
 we consider here the ansatz, 
\begin{equation}
\frac{v'} {v} = \xi'(r) \label{CON01}
\end{equation}
due to which the equation (\ref {Dsv14.1}) is written in the form 
\begin{equation}
 k r {M}_{,\ r} +[(k + 1) r v \xi'(r) + v]
{M}_{,\ v} = - 3 k M  \; .\label {Dsv14A.1}
\end{equation}
This equation has a general solution of the form,
\begin{equation}
M(r,v) = m_{o} \frac{e^{[3(k+1) \xi(r)]} } {v^{3k}}   \label{SoD15.1}
\end{equation}
where $m_{o} $  is an arbitrary positive constant,  
and $\xi(r)$ is a continuously differentiable function restrained by no-trapped surface condition 
at the beginning of the collapse and by the compatibility condition. 
The introduction of the equation
(\ref{CON01}) demands its compatibility with other 
field equations or their subsequent 
equation (\ref{RD18}), which is discussed in the Appendix.

${M}(r,v)$ expressed in equation (\ref{SoD15.1})  represents many classes of solutions 
 of equation (\ref {Dsv14.1})
but only those classes are physically realistic which satisfy
the energy conditions, which are regular and which
give $\rho \rightarrow \infty$ as $v \rightarrow 0$. 

Using equation (\ref{SoD15.1}), we write
\begin{eqnarray}
 && M_{0}(v)=M(0,v)= m_{o} \frac{e^{[3(k+1) \xi(0)]} } {v^{3k}},\nonumber \\  
 && M_{1}(v)= {M_{,r}(0,v)}=0  \label{M17} 
\end{eqnarray}
under the conditions  $\xi(0)=const.$, $\xi'(0)=0$.
Here $M_{1}(v)=0$ is in accordance with the requirement that 
the energy density has no cusps at the center. The density profile 
for such class of models takes the form,
\begin{equation}
\rho(r,v) = \frac{3 m_{o} e^{[3(k+1) \xi(r)]}}
{v^{3(k+1)}}.  \label{Sr16} 
\end{equation}

On integrating equation (\ref{CON01}), we obtain 
\begin{equation}
 v(t,r) = e^{\xi(r)} Q(t)  \label{V17}
\end{equation}
where $Q(t)$ is an arbitrary function due to integration.

We have used the function $v(t, r)$ as a catalyst to find solution of field equation  (\ref{Dsfe05}).
Hereafter, we use $(t, r)$ coordinates in understanding the collapse of the dense star.
The physical quantities in $(t, r)$ take the form
\begin{eqnarray}
&& R(t,r)= r v(t,r)= r e^{\xi(r)} Q(t)  ,  \label{R01} \\
&& F(t,r) = m_{o} \; r^3 e^{3\xi(r)} Q(t)^{-3k} , \label{F02} \\
&& \rho(t)= 3 m_{o} \;  Q(t)^{-3(k+1)}  \; .\label{RH02}
\end{eqnarray}
It is easy to verify that these equations together satisfy field equation (\ref{Dsfe05}).

At the dynamical equilibrium event 
$t=t_{i}$, $R(t_{i},r)= r e^{\xi(r)} Q(t_{i})= r_{1}$, and $ 0 < r_{1} < r_{b}$
 where $r_{b}$ is the radius of the collapsing cloud. 
The regular density distribution at the initial surface takes the form
\begin{equation}
\rho_{o}(t_{i}) = 3 m_{o} \;  Q(t_{i})^{-3(k+1)}. \label{Sr17}
\end{equation}
Since $\rho_{0} \geq 0$, we have
$m_{o} \geq 0 $ with $Q(t_{i})= \  const. $
Further, at some $t=t_{b}$,  $r=r_{b}$ 
which is the boundary of the cloud where pressure 
is zero, and where the interior is matched to a suitable
exterior metric, and this is discussed in section  IV.

The density profile of the collapsing cloud is homogeneous at all the epochs. 
Now since the particles are alike, as the collapse of the star evolves, the density  
increases relative to time only on subsequent time-hypersurfaces
and  this feature will be specified by $Q(t)$ through $ \dot{Q}(t) < 0$.
The collapse condition $\dot{R} < 0$ becomes $\dot{Q}(t) < 0$.

 At the singular epoch $t=t_{s}$, $Q(t)$ should converge to zero and so that
density would diverge as $t \rightarrow t_{s}$. 
So, next we aim to find such $Q(t)$ satisfying all the above conditions.

Since the density profile is homogeneous, on integrating equation (\ref {Dsfe06}), we obtain 
the metric function,
\begin{equation}
\nu(t, r) =  a(t). \label {Dcv14}
\end{equation}
and using time scaling freedom, we set $a(t) = 0$.

We know that the shell-crossing singularity is gravitationally weak and spacetime can be extended
through it 
\cite{PSJ}.
Therefore, we assume $R' > 0$,  hence 
from equation (\ref{Dsfe07}), we have 
\begin{equation}
G(t,r) = \textbf{d}(r) \label {Dg18}
\end{equation}
where $ \textbf{d}(r) $ is another arbitrary continuously
differentiable function of $r$. Further, we write
\begin{equation}
\textbf{d}(r) = 1 + r^2 \; \textbf{b}(r)  \label {Db17}
\end{equation}
where $\textbf{b}(r)$ is at least twice continuously differentiable. 

The metric takes the form,
\begin{equation}
 ds^2 = - dt^2 +  \frac {R'^{2} } { 1 + r^2 \; \textbf{b}(r)  } dr^2 + R^2(t,r) d\Omega^2 . \label {Dms01}
\end{equation}
In the study of Einsteins field equations with equation of state, the system of  equations gets closed
but  still, we have introduced equation (\ref{CON01}) so it needs its compatibility with the field equations.
It is found that for the case $\textbf{b}(r) = 0$,  the function $\xi(r)$ remains arbitrary in satisfying the compatibility condition. Now for such physically realistic function  $\xi(r)$,
 the mass profile of the collapsing cloud will be known in $r$ and thereafter,
solving constraint equation (\ref{Dsfe08}), we can obtain the function $Q(t)$,
 thus giving rise to an exact solution of the
system of equations.

 While for the case $\textbf{b}(r) \neq 0$, the choice of function $\xi(r)$
 is restricted by the condition $\textbf{b}(r) = \pm  b_{o} e^{2\xi(r)}$
where $b_{o}$ is a positive constant, thus 
shrinking the domain of the solution set. 

Now, in order to determine $Q(t)$, 
the field equation (\ref{Dsfe08}) can be expressed in the form
\begin{equation}
\dot{R}^2 =  \left[\frac{F}{R} + G -1 \right] \label{RD18}
\end{equation}
where $ G - 1= r^2 \textbf{b}(r)$, and so $\textbf{b}(r)$ basically characterizes 
the energy distribution for the collapsing shells.

Using equations (\ref{R01}) and (\ref{F02}), this can be written as,  
\begin{equation}
\dot{Q}(t) = -  e^{- \xi(r)} \sqrt{\frac{ m_{o} e^{2\xi(r)}}{Q(t)^{(1+3k)}}  +\textbf{b}(r) } \label{RD19}
\end{equation}
where the negative sign is chosen since for collapse
we must have $ \dot{Q} < 0 $.
Integration of  the above equation gives us,
\begin{equation}
 t(r,Q) = t_{i} + \int_{Q}^{Q(t_{i})}  \frac{  e^{\xi(r)} \; dQ}
 {\sqrt{\frac{  m_{o} e^{2\xi(r)}} {Q(t)^{(1+3k)}}  +\textbf{b}(r) }} 
 \label {Dsi21}
\end{equation}
where the variable $r$ is treated as a constant.
From equation  (\ref{Dsi21}), we have a regular time function at the center of the cloud
 \begin{eqnarray}
t(0,Q) = t_{i} + \int_{Q}^{Q(t_{i})}  \frac{  e^{ \xi(0)} \; dQ}
 {\sqrt{\frac{ m_{o} e^{2\xi(0)}} {Q(t)^{(1+3k)}}  +\textbf{b}(0) }} \; . \label{tov01}
\end{eqnarray}
The time for other collapsing shells to arrive at 
the singularity can be expressed by
\begin{equation}
 t_{s}(r)=t(r,0) = t_{i} +  \int_{0}^{Q(t_{i})}  \frac{  e^{ \xi(r)} \; dQ}
 {\sqrt{\frac{  m_{o} e^{2\xi(r)}} {Q(t)^{(1+3k)}}  +\textbf{b}(r) }} \; .
\label {Dt26}
\end{equation}
Since the energy density has no
cusps at the center, that means $M_{1} = 0$. The singularity 
curve then takes the form,
\begin{eqnarray}
t_{s}(r) = t_{0} + r \chi_{1}(0) + \frac{r^{2}}{2 !} 
\chi_{2}(0) + {\mathcal{O}}(r^{3}),  \label{TSR} \\
\text {where } \ 
\chi_{1}(Q) = \frac{dt}{dr} \Big{|}_{r=0} , \; \; 
\chi_{2}(Q) =  \frac{d^{2}t}{dr^{2}} \Big{|}_{r=0}  \nonumber 
\end{eqnarray}
and $t_{0} = t(0, 0)$ is the time at which the central 
shell becomes singular and which is obtained  as 
\begin {eqnarray}
 t(0,0)= t_{i} + \int_{0}^{Q(t_{i})}  \frac{  e^{ \xi(0)} \; dQ}
 {\sqrt{\frac{  m_{o} e^{2\xi(0)}} {Q(t)^{(1+3k)}}  +\textbf{b}(0) }} \; . \label{TOV}
\end{eqnarray}
The time taken by the central shell to reach the 
singularity should be positive and finite, and hence we have 
the {\it model realistic condition} (MRC), namely that,
{\small
\begin{equation}
\left[ \frac{  m_{o} e^{2\xi(0)}}   {Q(t)^{(1+3k)} } +\textbf{b}(0)  \right]> 0  \; \text{i.e.} \; 
Q(t)^{(1+3k)}  <  - \frac{  m_{o} e^{2\xi(0)}}{\textbf{b}(0)}  \label{MRC01}
\end{equation}
}
and it must be finite for any  $k \in (-1/3,1]$. 

From consistency of field equations, we have   
$\textbf{b}(0) = \pm b_{o} e^{ 2 \xi(0)} $ 
but at the same time $Q(t)$ is a positive real valued function, 
and above inequality makes this possible only when 
$ \textbf{b}(0) = - b_{o} e^{2 \xi(0)}$, giving rise to the range of $Q(t)$, 
$ 0 \leq Q(t) < [m_{o} /b_{o} ]^{1/(1+3k)}$ .  

 Thus the initial data of mass and density profiles is restricted by the 
introduction of the equation (\ref{CON01}) through the condition $ \textbf{b}(r) = - b_{o} e^{2 \xi(r)}$.

 For $\textbf{b}(0) = 0$, the above MRC takes the form $Q(t) \geq 0$. 
Now for various choices of $\xi(r)$,
we can study different models, for eg. in the case of $\textbf{b}(r)=0$
 if we choose $\xi(r)=0$,  metric (\ref{Dms01}) gives us Einstein-deSitter model with equation of state
while for $b_{o}=1, \xi(r)=0$, we have a closed Friedman model, and so on. \\

%%%%%%%%%%%%%%%%%
Now using $ \textbf{b}(r) = - b_{o} e^{2 \xi(r)}$,  equation (\ref{Dsi21}) takes the form 
\begin{equation}
 t(Q) = t_{i} + \int_{Q}^{Q(t_{i})}  \frac{ dQ}
 {\sqrt{\frac{  m_{o} } {Q(t)^{(1+3k)}}  - b_{0} }} .
 \label {Q27}
\end{equation}
On solving above equation, we have
\begin{eqnarray}
 t(Q) && = t_{i} + \frac{2  \Big[ Q(t_i)^{\frac{3 (k+1)}{2} }H_{1}
 - Q(t)^{\frac{3 (k+1)}{2} } H_{2}   \Big] } {3 \sqrt{m_{o} } (1+k) }
 \label {Q28} \\
\text{where} \; 
  H_{1} && = \text{hypergeom} ( [ 1/2, K_{1}] , [K_{2}],   b_{o} Q(t_i)^{K_{3}}/m_{o}  ), \nonumber \\
 H_{2} && = \text{hypergeom} ( [ 1/2, K_{1}] , [K_{2}],   b_{o} Q(t)^{K_{3}} /m_{o}  ) , \nonumber \\
K_{1} && = \frac{3(k+1)}{2(3k+1)}, K_{2}= \frac{(9k+5)}{2(3k+1)} \; \text{and} \; K_{3} = 3k+1. \nonumber 
\end{eqnarray}
%%%%%%%%%%%%
 The hypergeometric series mentioned above is convergent for $|b_{o} Q(t)^{K_{3}} /m_{o}| < 1$
 and $ -1/3 < k  \leq 1$. 
 The convergence condition on $ Q(t)$ augurs well with the MRC restriction
 $ 0 \leq Q(t) < [m_{o} /b_{o} ]^{1/(1+3k)}$ . \
We can find $\dot{Q}(t)$ from above equation
\begin{eqnarray}
 &&\dot{Q}(t)  = - \frac{2 {  m_{o} }^{3/2}  K_{2} Q(t)^{-(3k+1)/2}} 
{ [ 2 {  m_{o} }   K_{2} H_{2} + b_{o}  H_{3} Q(t)^{K_{3}}  ]}  \label {Qdot29} \\
&& \text{where} \nonumber \\
&& H_{3}  = \text{hypergeom} \left( [ 3/2, K_{1} + 1] , [K_{2} + 1],   b_{o} Q^{K_{3}} /m_{o}  \right) . \nonumber 
 \end{eqnarray} 
The collapse condition $ \dot{Q}(t) < 0$ for the dense cloud is thus satisfied and as $t \rightarrow t_{s}$,
 indicates perpetual gravitational collapse of the star.

The time taken by the central shell to reach the singularity is given by
\begin {eqnarray}
 t_{0}= t_{i} + \frac{2 Q(t_i)^{\frac{3 (k+1)}{2} } H_{1}} {3 \sqrt{m_{o}} (1+k) } .
\label{TOV1}
\end{eqnarray}
We observe through the implication of equation (\ref{Q28}) on equation (\ref{TSR}) is
that $t_{s}(r) = t_{0}  $,   
indicating that time of formation of central singularity $ (t=t_{s}, r=0)$ and 
the non-central singularity $ (t=t_{s}, r= r_{c} > 0) $ in the neighbourhood 
of the center $r =0$ is same. Clearly these events are simultaneous and it is expected  that
in such scenario the singularity be remain covered behind the event horizon.
%%%%%%%%%%%%%%%%%%%%%

In the analysis of gravitational collapse of a dense star, 
we must have the initial configuration to be not trapped.
Therefore,  we must have
\begin{equation}
\frac{F(t_{i},r)} {R(t_{i},r) } =  m_{o} r^2 e^{2 \xi(r)}  Q(t_{i})^{-(1+3k)}  < 1 \;, \label{FRT}
\end{equation}
this allows for the formation of trapped surfaces during the collapse. 
This constraint indicates, 
how the choice of the initial matter configuration \textbf{$F(t_{i},r)$} through $\xi(r)$ is related
to the initial surface of the collapsing cloud. Some restrictions on the choices of function $ \xi(r)$
must be made in order, not to have trapped surfaces at the
initial time.  Further, to prevent trapped surfaces at the initial epoch the velocity of the
infalling shells must satisfy  
$|\dot{R} | > \sqrt{|d(r)|}  .$  
It clearly shows  that the initial velocity of the infalling shells of the cloud must always be positive and that the
case of equilibrium configuration where $\dot{R} = 0$ can be
 taken only at the static boundary of the star where pressure is zero
\cite{DPJRVS}.

\section{The apparent horizon}

When the singularity curve is
constant ($\chi_1$ and other higher order terms are all vanishing),
or would be decreasing, then a black hole will necessarily
form as the collapse final state. For a black hole to come into being the trapped surfaces
form before the formation of singularity. Thus for a black hole to
form we require, \cite{SSJ}
\begin{equation} 
t_{\rm ah}(r)\leq t_0~\mbox{for}~r>0\,,~\mbox{near}~r=0  \label{bh}
\end{equation}
where $t_0$ is the epoch at which the central shell
hits the singularity.

We know since the collapsing shells are simultaneous, the end state of collapse is bound to be a
black hole or to say condition (\ref{bh}) will be satisfied but the intriguing question is that,
what is the role of parameter $k$ of equation of state in the formation of the black hole.
Is their a certain range of $k$
in which the formation of black hole will be accelerated in time?

The introduction of equation (\ref{CON01}) has imposed restriction on the function 
$\textbf{b}(r)$ through compatibility of field equations and that now we have $\textbf{b}(r)= - b_{o} e^{\xi(r)}$. Therefore, we have 
only two cases to study namely that spacetime is bound or marginally bound.
The time of occurrence of apparent horizon in a bounded spacetime is written as
\begin{equation}
 t_{bah} = t_{bs} - \int_{0}^{Q_{ah}}  \frac{ dQ}
 {\sqrt{\frac{  m_{o} } {Q(t)^{(1+3k)}}  - b_{0} }} \  
 \label {B01}
\end{equation}
where $t_{bs} = t_{0}$, given by equation (\ref{TOV1}) and $Q(t_{ah}) \equiv Q_{ah}$
 obtained from $F/R=1$  is specified as
\begin{equation}
 Q_{ah}  = \left[m_{o} \; {r_{ah}}^2 \; e^{2 \xi(r_{ah})}   \right]^{\frac{1} {(1+3k)}} \; , 
 \label {B02}
\end{equation}
further we can express
\begin{equation}
R_{ah}  = \frac{1}{\sqrt{m_{o}} } \left[m_{o} \; {r_{ah}}^2 \; e^{2 \xi(r_{ah})}   \right]^\frac{3(1+k)}{[2(1+3k)]}.
 \label{Rah}
\end{equation}
On solving equation (\ref{B01}), we have 
\begin{eqnarray}
 t_{bah} && = t_{bs} - t_{bk}  \  \ \text{where} \ \
  t_{bk}  =  \frac{2 Q_{ah}^{\frac{3 (k+1)}{2} } H_{2ah}} {3 \sqrt{m_{o} } (1+k) }    
   \label {B03} \\
  \text{and} \; H_{2ah} && = \text{hypergeom} ( [ 1/2, K_{1}] , [K_{2}],   b_{o} Q_{ah}^{K_{3}}/m_{o}  ).
 \nonumber 
\end{eqnarray}
It is clear that $t_{bk}$ is a positive quantity for all $k \in (-1/3, 1]$, and therefore 
$t_{bah} < t_{bs}$ for any $r > 0$, near the center $r = 0$. 
The collapse progresses to culminate into the formation of trapped surfaces first
and eventually the singularity forms later,
 leading to formation of BH as a final state of collapse for all $k \in (-1/3, 1]$.

Now, we study the characteristics of the parameter $k$ in the formation of the BH. 
Let us consider the situation where we have known initial mass and physical radius
of the collapsing star, and this leads to the determination of component $m_{o}$ of mass or density of the star. 
And since our model is homogeneous in density, therefore this density shall be alike
at all $r$ at the initial epoch. 

Using equations (\ref{R01}) and (\ref{F02}) we can write 
\begin{equation}
m_{o} = \frac{ F(t_{i}, r)  Q(t_{i}) ^{3(1+k)} } { R^{3}(t_{i}, r) }  \label{mo}
\end{equation} 
and we find that $m_{o}$ decreases as $ k \rightarrow 1$ because 
$ 0 \leq Q(t) < [m_{o}/b_{o}]^{1/(1+3k)} < 1$.  
We can rearrange the expression for $t_{bk}$ given in equation (\ref{B03})  as follows
\begin{eqnarray}
   t_{bk} && =  \frac{2 \; R_{ah} \; H_{2rah} } 	{3 (1+k) }   \label {B033} \\
		\text{where} \; 
  	 H_{2rah} && =\text{hypergeom} ( [ 1/2, K_{1}] , [K_{2}], z  )   \nonumber \\
		\text{and} \; z && =  b_{o} \; {r_{ah}}^2 \; e^{2 \xi(r_{ah})} . \nonumber
  \end{eqnarray}
	
Now the star with a known radius and mass collapses under its own gravitational pull
towards the center which is solely caused due to its mass, and not because of what composition of matter it has. 
The boundary of the horizon of such a star can be known through $ F = R$, this gives us 
a specific radius $Rah$, independent of the equation of state
but equation of state can stimulate the scenario of formation of trapped surfaces. It is indeed possible to testify
whether formation of trapped surfaces of such a star would accelerate or decelerate in time
relative to change in parameter of equation of state. In view of these aspects the theorem follows: \\
%%%%%%%%%%%%%%%%%%%%%
{\bf Theorem 1}. \\
Consider $t_{bk} = t_{bk}(k, r_{ah})$,   $r_{ah}$ depends on $k$  and
$ 0 < Q_{ah} < Q(t_{i}) < 1$.  We prove that both $t_{bk}$ and $t_{bs}$ are positive decreasing
 time functions as $k \rightarrow 1$ and that $t_{bs} > t_{bk}$ for all $k \in (- 1/3, 1]$. 
Further  $t_{bah} < t_{bs}$ for any $r > 0$, near the center $r = 0$ and 
 $t_{bh}$ is a positive decreasing  time function as $k \rightarrow 1$ .\\

{\bf Proof}: Since $R_{ah}$ remains same for the given mass and physical radius
 of the collapsing star, irrespective of the different values of equation of state parameter $k$, 
then equation (\ref{Rah}) dictates $m_{o}$ and $r_{ah}$ 
to vary relative to $k$.
Therefore using equations  (\ref{Rah}), (\ref{mo}) and (\ref{B033}), we have  
\begin{eqnarray}
&& \frac{dr_{ah}} {dk} = - \frac{ r_{ah} \; ln\left[ \frac{Q_{ti}} {Q_{ah}}\right]  } 
{ (1+k) [1 +  r_{ah} \xi'(r_{ah})] }   \;  < 0 \label{TH02} \\
&& \frac{\partial t_{bk}} {\partial r_{ah}} = \frac{ 2 b_{o} R_{ah} 
   r_{ah} \; e^{2 \xi(r_{ah})}  H_{3rah}  } {(9k+5) [1 +  r_{ah} \xi'(r_{ah})]^{-1}} \; > 0
\label{TH03} \\
&& \frac{\partial t_{bk}} {\partial k} = \frac{ 2 \; R_{ah}} {3 (1+k)^2 } 
\left[(1+k) \frac{\partial}{\partial k} H_{2rah} -  H_{2rah} \right] \; < 0
\label{TH04} \\
&&\text{where} \; H_{3rah}  = \text{hypergeom} ( [ 3/2, K_{1} +1] , [K_{2}+1],  z  ) 
 \nonumber 
\end{eqnarray}
and signs are prescribed under the conditions that $ 0 < Q_{ah} < Q(t_{i}) < 1$,     $Q(t_{i})/ Q_{ah} > 1$ and $|z|<1$.
These physically realistic conditions are possible with the appropriate choice of the function $\xi(r)$
such as $[1 +  r_{ah} \xi'(r_{ah})]  > 0$.
Now, we can write
\begin{equation}
 \frac{dt_{bk}} {dk} =  \frac{\partial t_{bk}} {\partial k} 
 + \frac{\partial t_{bk}} {\partial r_{ah}} \frac{dr_{ah}} {dk} \; \; < 0.\label{TH05}
\end{equation}
We can obtain $dt_{bs}/dk$  using equations (\ref{TOV1}) and (\ref{mo}), 
\begin{eqnarray}
\frac{dt_{bs}} {dk} && = \frac{2 Q(t_{i})^{3(k+1)/2 }} { 3 (1 +k )^2 \sqrt{m_{o}} }
 \left[(1+k) \frac{\partial}{\partial k} H_{1} - H_{1}   \right]  \;  < 0 \hspace{0.1in} \label{TH06}  
\end{eqnarray}
From equations (\ref{TOV1}) and (\ref{B03}), we have $t_{bs} > 0$ and $t_{bk} > 0$, further  we can write
\begin{equation}
\frac{t_{bs}} { t_{bk}} = \frac{ 3 \sqrt{m_{o}} (1+k) \; t_{i}  } { 2 \; H_{2ah} \; Q_{ah}^{3(1+k)/2} }
 +\left[\frac{H_{1}} {H_{2ah}} \right] \left[ \frac{Q(t_{i})} {Q_{ah}} \right]^{\frac{3(1+k) } {2} }  > 1 
\label{TH07}  
\end{equation}
%%%%%%%%%%%
Therefore, from above we conclude that
  $t_{bk}$ and $t_{bs}$ are positive decreasing time functions as $k \rightarrow 1$. \\
	
Now $t_{bah} = t_{bs} - t_{bk}$ \ and \ $ t_{bs} > t_{bk} > 0$, therefore
$t_{bah} < t_{bs}$ for any $r > 0$, near the center $r = 0$.
 Clearly indicating that  trapped surfaces are being formed first, 
and the event of the formation of singularity is taking place at the later time.
Thus black hole forms for all $k$.

 Further since  both $ t_{bs}$ and $t_{bk}$ are positive decreasing functions as 
$k \rightarrow 1$ and  $ t_{bs} > t_{bk}$. Therefore
 $t_{bah}$ is a positive decreasing function as $k \rightarrow 1$.  \  $\bf{\diamondsuit}$  \\

%%%%%%%%%%%%%%%%%%%%%%%%
In the marginally bound case that is when  $\textbf{b}(r)=0$, on integrating equation (\ref{RD19}), we have 
\begin{equation}
Q(t) =  \left[ \frac{3} {2} \sqrt{ m_{o}} (1+k) (t_s - t)  \right]^{2/[3 (1+k)] }  \label{MBQ01}
\end{equation}
then $\dot{Q}(t) < 0$ and as $t \rightarrow t_{s}$, $\dot{Q}(t) \rightarrow - \infty$.
The physical radius and density of the collapsing star are obtained as
\begin{eqnarray}
&& R(t,r) =  r e^{\xi(r)} \left[ \frac{3} {2} \sqrt{ m_{o}} (1+k) (t_s - t)  \right]^{2/[3 (1+k)] } \nonumber \\
&& \rho(t) = \frac{4}{3 (1+k)^2 (t_s - t)^2 }  \;  .\label{MBQ02}
\end{eqnarray}
The apparent horizon equation using equation (\ref{MBQ01}) is expressed by
\begin{eqnarray}
&& t_{ah} = t_s - t_k \;   \;  \text{where} \nonumber \\ 
&& t_k =  \frac{2 \left[m_{o} \; {r_{ah}}^2 \; e^{2 \xi(r_{ah})}  \right]^{ \frac{3(1+k)} {(2+6k)} } } {3(1+k)\sqrt{m_{o} }} 
 =  \frac{2 R_{ah} } {3(1+k)} \ .
\label{MBQ03}
\end{eqnarray}
Time taken by the shells to reach the singularity is given by equation (\ref{TOV1})
 with  $H_{1}=1$ ( or by equation (\ref{MBQ01}) )
\begin{equation}
t_{s} = t(0) = t_i + \frac{2} {3 (1+k) \sqrt{m_{o}}  } Q(t_i)^{3(1+k)/2} .
\label{MBQ04}
\end{equation}
{\it Clearly, Theorem 1. holds for marginally bound space-time wherein $b_{o} =0$. } \\
%%%%%%%%%%%%%%%%%%%%
 It is evident from Theorem 1. that the equation of state is stimulating the formation of
apparent horizon of gravity  to take place
 at the earlier epoch and further strengthening this characteristic as $k$ increases
as compared to the usual process of formation of trapped surfaces
in the final stages of collapse of the sufficiently large star,  culminating it into the 
black hole at the earlier time. This process is accelerated in time as $k \rightarrow 1$
 with  the physically plausible choice of the function $\xi(r)$.

To have further insight into the end stages of collapse of the star and to analyze the role of parameter $k$, 
we consider an example of a Neutron star temporarily in equilibrium state, 
having a mass of  $3.24 M_{ \odot}$ and physical radius of
$18.02 Km$.  This star in due course of time collapses under its gravitational force.
The dominant tidal force culminates it into the BH with mass of  $1.7192 M_{ \odot}$
and the radius shrinks to $5.07 Km$ where mass $F=2 \ \texttt{M}$
\cite{ST}. 
We have at the horizon,
  $R_{ah}$ given by equation (\ref{Rah}), thereby
$r_{ah}$ is analyzed through numerical solution by expanding $e^{ 2\xi(r_{ah})}$ to the
 second order, 
 for the physically plausible choice of the function $\xi(r)$. 
The results are shown in Table I \& II.
%%%%%%%%%%%%%%
\begin{table}
\label{Table -01}
\begin{center}
\caption{Role of $k$ is analyzed through the time of formation of horizon and that of singularity 
for initial mass of $3.24 M_{ \odot}$  and physical radius of $18.02 Km$. 
The collapse end state BH has mass of  $1.7192 M_{ \odot}$
and the Schwarzschild radius is of $5.07 Km$ 
with $\xi(r) = 1 + r^2/2 $, $t_i =0$, $Q(t_i)=0.37$  
and  $b_{0}=0.2227 \times 10^{-3}$ . \label{tbl-1}}

\begin{tabular}{lcccccc}
\hline 
\hline

$k$ & $ m_{o}  $ &$  r_{ah} $ & $z$ & $t_s $ & $t_{bk}$ & $t_{bah}= t_s - t_{bk}$ \\
\hline

$-0.3$  &$0.20\times 10^{-3}$ &  $4.216$     &$0.856$      & $109.33$        & $10.65$      & $98.6848$ \\
$-0.2$   &$0.15\times 10^{-3}$ & $3.930$     &$0.587$      & $58.50$					& $5.73$				 & $52.7684$   \\
$-0.1$   &$0.11\times 10^{-3}$ & $3.719$     &$0.438$      & $43.74$					& $4.47$				 & $39.2577$   \\
$0$       &$0.83\times 10^{-4}$  & $3.556$    &$0.346$      & $35.79 $         & $3.82$         & $31.9723$ \\
$0.1$    &$0.61\times 10^{-4} $ & $3.427$    &$0.286$      & $30.59$         & $3.36$        & $27.2268$ \\
$0.2$    &$0.46\times 10^{-4}$  & $3.322$    &$0.243$      & $26.85$          & $3.03$         & $23.8244$ \\
$0.3$    &$0.34\times 10^{-4}$  & $3.235$    &$0.213$      & $23.99$         & $2.76$        & $21.2377$ \\
$0.4$    &$0.25\times 10^{-4}$  & $3.162$    &$0.189$     & $21.73 $         & $2.54$        & $19.1912$ \\
$0.5$    &$0.19\times 10^{-4}$  & $3.099$    &$0.171$      & $19.88$          & $2.35$          & $17.5245$ \\
$0.6$   &$0.14\times 10^{-4}$   &$3.045$     &$0.157$      & $18.33$         & $2.20$        & $16.1370$ \\
$0.7$    &$0.10\times 10^{-4}$  & $2.998$    &$0.145$      & $17.02$         & $2.06$        & $14.9614$ \\
$0.8$  &$0.76\times 10^{-5}$    & $2.957$    &$0.135$      & $15.89 $         & $1.94$        & $13.9512$ \\
$0.9$   &$0.56\times 10^{-5}$   & $2.920$    &$0.127$      & $14.90$         & $1.83$        & $13.0728$ \\
$1.0$   &$0.42\times 10^{-5}$   & $2.887$    &$0.120$      & $14.04$         & $1.74$        & $12.3014$ \\

\hline 
\hline
\end{tabular} 
\end{center}
\end{table}
%%%%%%%%%%%%%%%%%%%%

%%%%%%%%%%%%%%%%%%%
\begin{table}
\label{Table -02}
\begin{center}
\caption{Time of formation of singularity and horizon are studied
 for the same data conceived in Table I except that  $b_{o}=0$. Herein $z =0$. }

\begin{tabular}{lccccc}
\hline
\hline

$k$ & $ m_{o}  $ &$  r_{ah} $  & $t_s $ & $t_{k}$ & $t_{ah}= t_s - t_{k}$ \\
\hline

$-0.3$  &$0.20\times 10^{-3}$   &  $4.216$     & $23.5566$    & $4.8286$    & $18.7280 $ \\
$-0.2$   &$0.15\times 10^{-3}$  & $3.930$      & $20.6120$					& $4.2250$				 & $16.3870$   \\
$-0.1$   &$0.11\times 10^{-3}$  & $3.7186$    & $18.3218$					& $3.7556$				  & $14.5662$   \\
$0$       &$0.82\times 10^{-4}$  & $3.5559$    & $16.4896$         & $3.3800$         & $13.1096$ \\
$0.1$    &$0.61\times 10^{-4} $  & $3.4268$   & $14.9905$         & $3.0727$         & $11.9178$ \\
$0.2$    &$0.46\times 10^{-4}$  & $3.3219$    & $13.7413$          & $2.8167$         & $10.9247$ \\
$0.3$    &$0.34\times 10^{-4}$  & $3.2350$    & $12.6843$         & $2.6000$          & $10.0843$ \\
$0.4$    &$0.25\times 10^{-4}$  & $3.1619$    & $11.7783$         & $2.4143$          & $9.3640$ \\
$0.5$    &$0.19\times 10^{-4}$  & $3.0994$    & $10.9931$          & $2.2533$          & $8.7397$ \\
$0.6$   &$0.14\times 10^{-4}$   & $3.0455$    & $10.3060$         & $2.1125$         & $8.1935$ \\
$0.7$    &$0.10\times 10^{-4}$  & $2.9984$    & $9.6998 $         & $1.9882$        & $7.7115$ \\
$0.8$  &$0.76\times 10^{-5}$    & $2.9570$    & $9.1609$         & $1.8778$        & $7.2831$ \\
$0.9$   &$0.57\times 10^{-5}$  & $2.9203$     & $8.6787$         & $1.7789$        & $6.8998$ \\
$1.0$   &$0.42\times 10^{-5}$   & $2.8876$    & $8.2448$         & $1.6900$        & $6.5548$ \\

\hline
\hline
\end{tabular} 
\end{center}
\end{table}
%%%%%%%%%%%%%%%%%%%%%%

In Table I and II, we have observed that as $ k \rightarrow 1$, 
the time of formation of singularity $t_s$ decreases in time,
 and that the time of formation of event horizon of gravity $t_{ah}$ as well decreases 
 but precedes in time to $t_s$. So the trapped surfaces form well in advance in time before
 the event of formation of singularity takes place, culminating the final stages of collapse into a black hole. 
Further the formation of BH is accelerated in time as $ k \rightarrow 1$ 
in the sense that trapped surfaces are coming into existence at the earliest time
and thus this property is strengthened for increasing parameter of equation of state.
%%%%%%%%%%%%%%%%%%%%%%%%%%%%%%%%%%%%
\section {Exterior space-time and junction conditions}
The space time in the exterior region denoted as ${\mathcal M_{+}}$ of the collapsing stellar configuration will be filled with radiation flowing outward along radial direction and it is appropriately described by the Vaidya metric 
\begin{equation}
ds^2_{+} = - \left(1 -  \frac{2 M(V)}{{y}} \right) dV^2 -
        2 \; dV d{y} + {y}^2 d \Omega^2   . \label{eq:iem14}
\end{equation}

Let ${\mathcal M_{-}}$ denote the space time in the interior of the collapsing star which is separated from the exterior by a time like 3 dimensional space time surface $\Sigma$ which represents at any instant the boundary separating  ${\mathcal
M_{+}}$ from  ${\mathcal M_{-}}$. The intrinsic metric on $\Sigma$ will be
\begin{equation}
ds^2 =  - d\tau^2 + {\mathcal R}^2(\tau) d\Omega^2 \label{eq:bmJ03}
\end{equation}
The metric on the interior manifold $\mathcal M_{-}$ is described by 
\begin{equation}
 ds^2_{-} = - dt^2 +  \frac {R'^{2} } { 1 + r^2 \; \textbf{b}(r)  } dr^2 + R^2(t,r) d\Omega^2 . \label {DmsJ04}
\end{equation}

 The boundary conditions smoothly joining the interior and exterior manifold
 $\mathcal M_{-}$ and $\mathcal M_{+}$ across $\Sigma$ are stipulated as \cite{GD,isa}, 
\begin{equation}
(ds_{-}^2)_{\Sigma} = (ds_{+}^2)_{\Sigma} = (ds^2)_{\Sigma}
\label{eq:jc18}
\end{equation}
%%%
\begin{equation}
K^-_{ij} = K^+_{ij} \label {eq:jc19}
\end{equation}
where 
\begin{equation}
K^{\pm}_{ij} = - n_{\alpha}^{\pm} \frac{\partial^2
x^{\alpha}_{\pm}}{\partial \xi^i \partial \xi^j} -
n_{\alpha}^{\pm} \Gamma^{\alpha}_{\beta \gamma} \frac{\partial
x^{\beta}_{\pm} }{\partial \xi^i} \frac{\partial x^{\gamma}_{\pm}
}{\partial \xi^j}
 \label{eq:jc20}
\end{equation}
denote extrinsic curvatures of $\Sigma$ in $\mathcal M_{\pm}$ respectively \cite{ST,GSS}.

The corresponding normal vectors are 
 \begin{equation}
n_{\alpha}^- = \left(0, \frac{R'} {\sqrt{1 + r^2 \textbf{b}(r)}},0,0  \right) ,
n_{\alpha}^+ = \frac{ dV}{d\tau} (- \frac{d{y}}{dV},1,0,0) .
\label{eq:un21}
\end{equation}
The  boundary conditions (\ref{eq:jc18}) imply the following relations
\begin{equation}
\frac{dt}{d\tau} =1 \; , \;  R(t, r) = {\mathcal R}(\tau) = {y} 
\label{eq:mc23}
\end{equation}
\begin{equation}
\text{and} \hspace{0.2in} \left( \frac{dV}{d\tau} \right)^{-2}_{\Sigma} = \left(1 -
\frac{2 M(V)}{y} + 2 \frac{d{y}}{dV} \right)_{\Sigma} \; . 
 \label{eq:jc24}
\end{equation}
The extrinsic curvatures $K_{ij}$ of $\Sigma$ are found to have the following explicit expressions
\begin{subequations}
%%%%%
\begin{eqnarray}
&& K^{-}_{\tau \tau} = 0_{\Sigma}
\label{eq:ec25a} \\
&& K^{+}_{\tau \tau} = \left[ \frac{d^2V}{d\tau^2} \left(\frac{dV}{d\tau} \right)^{-1}
- {\frac{M(V)}{{y}^2} } \frac{dV}{d\tau}\right]_{\Sigma}
  \label{eq:ec25c} \\
&& K^{-}_{\theta \theta}  = \left[ R \; \sqrt{ 1 + r^2 \textbf{b}(r) } \right]_{\Sigma}
  \label{eq:ec25b} \\
&& K^{+}_{\theta \theta} = \left[ {y} \frac{d{y}}{d \tau}
 + {y} \frac{dV}{d\tau}  \left(1 - \frac{2 M}{y} \right)
 \right]_{\Sigma} \label{eq:ec25d}  \\
%%&& K^{\pm}_{\phi \phi} = \sin^2 \theta K^{\pm}_{\theta \theta}  \label{eq:ec25e}  \\
&& K^{-}_{ij} = K^{+}_{ij} = 0 \; \text{for} \; i \neq j
\label{eq:ec25f}
\end{eqnarray}
%%%%
\end{subequations}
In view of Eqs. (\ref{eq:ec25a}) to (\ref{eq:ec25d}), the boundary conditions ensuring continuity of extrinsic curvatures across $\Sigma$ imply the following relations
\begin{equation}
 \left[ \frac{d^2V}{d\tau^2}  \right]_{\Sigma} = 
\left[ \frac{M(V)}{{y}^2} \left( \frac{dV}{d\tau} \right)^2 \right]_{\Sigma} 
 \label{eq:mf31}
\end{equation}
%%%%%%%%
\begin{equation}
\left[ {y}\frac{d{y}}{d\tau} + {y} \frac{dV}{d\tau}
\left(1 - \frac{2M(V)}{y}  \right)
\right]_{\Sigma} =\left[ R \sqrt{ 1 + r^2 \textbf{b}(r) } \right]_{\Sigma}.   \label{eq:mf26}
\end{equation}
%%%%%%
The equations (\ref{eq:mc23}) and (\ref{eq:mf26}) determine the mass 
 contained within spherical region in ${\mathcal M_{+}}$ as 
\begin{equation}
M(V) = \left[ \frac{ R}{2} \left( \dot{R}^2 - r^2 \textbf{b}(r) \right)
 \right]_{\Sigma} \label{eq:emf27}
\end{equation}
%%%
%%%%%%%%%%%%%%%%%%%%%%%%%%%%
  We obtain $dV/{d\tau}$ and $d^2V/{d \tau^2}$ from equations (\ref{eq:mc23}) and (\ref{eq:jc24}),
	on using them together with equation (\ref{eq:emf27}), we find that the condition (\ref{eq:mf31})
	leads to the relation
	\begin{equation}
	 \left[2 R \ddot{R}  \right]_{\Sigma} = \left[r^2 \textbf{b}(r) - \dot{R}^2 \right]_{\Sigma}
	= - \left[\frac{F} {R} \right]_{\Sigma} \; \; . \label{J11}
	\end{equation}
	Further, on use of the equations (\ref{R01}) and (\ref{RD18}) , we can write  
	\begin{eqnarray}
	\left[(1 + 3 k)  \; m_{o} \; r e^{\xi(r)} Q(t)^{- (2+3k)}  \right]_{\Sigma}
	= \left[ \frac{F} {R^2}  \right]_{\Sigma} \; .\label{J12}
	\end{eqnarray}
	Finally using equation of state $p = k\rho$ together with equations (\ref{F02}) and (\ref{RH02}), we obtain
\begin{eqnarray}
  \left[p \;  Q(t) \right]_{\Sigma} = 0.
  \label{eq:pb32}
\end{eqnarray}
We know at the boundary of the star, the mass and the physical  radius  of the star are fixed numbers,  
and hence at the boundary  $Q(t)_{\Sigma} \neq 0$. 
Hence we must have 
\[ \left(p  \right)_{\Sigma} = 0.  \] 
Thus we could establish a complete star model which initially match with radiating Vaidya region,
and thereof with
the empty exterior represented by Schwarzschild metric.

%%%%%%%%%%%%%%
\section{Conclusions and Remarks}
Let us summarize the results, firstly we have obtained the solution of Type I matter field equations
through the ansatz introduced in equation (\ref{CON01}). Certainly, this has led to a special class of solutions
with an isentropic equation of state $p = k \rho$ that satisfy weak energy conditions and
evolve as the collapse begins according to the homogeneous distribution of matter.  

With the varied choices of function $\xi(r)$ satisfying physically realistic conditions,
 we have a class of bound and marginally bound space-times which can be explored further.
It is shown that how the choice of initial data of mass function and the physical radius through 
the function $\xi(r)$ lead to the formation of BH.

Studies show that gravitational lensing is an 
important astrophysical tool to observationally test the Cosmic Censorship Hypothesis (CCH) \cite{VE,SSMPNPJ}. 
Relativistic images of Schwarzschild black hole lensing is studied by K. S. Virbhadra \cite{KSV}.
In view of these aspects, we emphasize that the BH model presented here
may serve as an example to understand black hole physics in the light of proving or 
formulating CCH in dynamical gravitational collapse.

The investigation of gravitational collapse with a linear 
equation of state has revealed the role of the parameter $k$ in
terms of formation of BH and further strengthening it by accelerating 
the formation of trapped surfaces in time, in both the bound and marginally bound space-times.

The parameter value $k =1$ depicts the case of stiff fluid ( 
that the equation of state becomes rigid enough )
 which itself may halt the progress of the collapse at some stage
\cite{VPZ}. 
Therefore our results are more significant in the range of $-1/3 < k < 1$. 

 Though this aspect together with whether 
the equation of state would remain unchanged or it would 
actually evolve and change as the collapse develops could not be established
because of complexities. Also the unbound case of space-time 
could not be studied with the solution exhibited because of restrictions imposed
by the compatibility conditions.

{\bf{Acknowledgement}}: \\
Sanjay Sarwe acknowledges the facilities extended by IUCAA, Pune, India
where part of this work was  completed under its Visiting Research Associateship Programme.

%%%%%%%%%%%%%%%%%
\section*{Appendix-\ Compatibility condition}
The introduction of the equation
(\ref{CON01}) which is given by 
$v'(t,r)= v(t, r)  \xi'(r)$, demands its compatibility with other field equations or their subsequent 
equation (\ref{RD19}). So, we consider
\begin{equation}
v' =  W(t,r,v, v', \dot{v}) \; \text{and}  \; \dot{v} = U(t,r,v, v', \dot{v}) . \label{ACC-1}
\end{equation}

 The condition of compatibility for non-linear partial differential equations of order one yields,
\begin{equation}
W_{,t}  =  U_{,r} \; . \label{AC-01}
\end{equation}

We have $W =  v(t,r) \xi'(r)$ and $ U = - \sqrt{D(t,r)}$, 
above equation takes the form 
\begin{eqnarray}
 &&  - \sqrt{D(t,r)} \; \xi'(r)    = - \frac{D'(t,r)}{2 \; \sqrt{D(t,r)}}     \nonumber \\
 \text{where} \;\; && D(t, r) = \frac{m_{o} \; e^{2 \xi(r)} }{ Q(t)^{1+3k}} + \textbf{b}(r).  \label{AC-02}
\end{eqnarray}
On simplification, we obtain
\begin{equation}
\frac{db}{dr} = 2 \textbf{b}(r) \frac{d \xi}{dr} \label{AC-03}
\end{equation}
and solving this equation, we have requisite condition of compatibility, 
\[ \textbf{b}(r) = \pm \  b_{o} e^{2 \xi(r)} . \]
\end {document}